\documentclass[12pt,onecolumn, draftclsnofoot]{IEEEtran}

\usepackage[subpreambles=true]{standalone}
\usepackage[reqno]{amsmath}
\usepackage{amssymb}

\usepackage{algorithm}
\usepackage{algorithmicx}
\usepackage{algpseudocode}
\usepackage{setspace}
\usepackage{multicol}
\usepackage{enumerate}
\usepackage[font=small, belowskip=-15pt,aboveskip=0pt]{caption}
\usepackage{cite}
\usepackage{amsthm}
\usepackage{cleveref}

\usepackage[]{graphicx}
\usepackage{tikz}
\usepackage{tkz-tab}
\usepackage{pgfplots}
\usepackage{booktabs}
\usepackage{floatflt}

\usepackage{psfrag}	
\usepackage{array}
\usepackage{multirow,hhline}
\usepackage{exscale}
\usepackage{color}
\usepackage{colortbl}
\usepackage{epsfig}
\usepackage{epstopdf}

\usepackage{array}
\usepackage[tight]{subfigure}

\usepackage[latin1]{inputenc} 
\usepackage[T1]{fontenc} 
\usepackage{rotating}
\usepackage{pbox}

\usepackage{pifont}

\usepackage{comment}
\usepackage[normalem]{ulem}
\usepackage{cancel}

\usepackage{bigints}

\newcommand{\mat}[1]{\ensuremath{\boldsymbol{#1}}}
\renewcommand{\vec}[1]{\ensuremath{\boldsymbol{#1}}}

\newtheorem{remark}{Remark}

\usetikzlibrary{shapes,snakes}
\usetikzlibrary{calc}
\usetikzlibrary{patterns}
\usetikzlibrary{decorations.pathmorphing} 
\usetikzlibrary{spy}
\usetikzlibrary{positioning}
\usetikzlibrary{arrows, decorations.markings}
\usetikzlibrary{shapes.multipart, shapes.geometric, fit, backgrounds}
\usetikzlibrary{spy}

\definecolor{mycolor1}{rgb}{0.00000,0.44700,0.74100}%
\definecolor{mycolor2}{rgb}{0.85000,0.32500,0.09800}%
\definecolor{mycolor4}{rgb}{0.92900,0.69400,0.12500}%
\definecolor{mycolor3}{rgb}{0.49400,0.18400,0.55600}%
\definecolor{mycolor5}{rgb}{0.46600,0.67400,0.18800}%
\definecolor{mycolor6}{rgb}{0.30100,0.74500,0.93300}%
\definecolor{mycolor7}{rgb}{0.63500,0.07800,0.18400}%

\begin{document}

\sloppy

\title{Event-Based Beam Tracking with Dynamic Beamwidth Adaptation in Terahertz (THz) Communications}
\author{\IEEEauthorblockN{Yasemin Karacora\IEEEauthorrefmark{1}, Christina Chaccour\IEEEauthorrefmark{2}, Aydin Sezgin\IEEEauthorrefmark{1}, and Walid Saad\IEEEauthorrefmark{2}\\}
    \IEEEauthorblockA{\IEEEauthorrefmark{1}{\small Institute of Digital Communication Systems, Ruhr University Bochum, Germany,\\\IEEEauthorrefmark{2}Bradley Department of Electrical and Computer Engineering, Virginia Tech., USA\\\vspace{-0.3cm} Emails: \{yasemin.karacora, aydin.sezgin\}@rub.de, \{christinac, walids\}@vt.edu}}
\thanks{This research was supported, in part, by the Federal Ministry of Education and Research (BMBF) of the Federal Republic of Germany under Grant 16KISK037 (6GEM), and by the U.S. National Science Foundation under Grant CNS-1836802.
}}

\maketitle
\vspace{-1.6cm}
\begin{abstract}
Terahertz (THz) communication will be a key enabler for next-generation wireless systems.
While THz frequency bands provide abundant bandwidth and extremely high data rates, their effective operation is inhibited by short communication ranges and narrow beams, thus, leading to major challenges pertaining to user mobility, beam alignment, and handover. In particular, there is a strong need for novel beam tracking methods that consider the tradeoff between enhancing the received signal strength via increasing beam directivity, and increasing the coverage probability by widening the beam. 
In this paper, a multi-objective optimization problem is formulated with the goal of jointly maximizing the expected rate and minimizing the outage probability subject to transmit power and overhead constraints.
Subsequently, a novel parameterized beamformer with dynamic beamwidth adaptation is proposed. 
In addition to the precoder, an event-based beam tracking approach is introduced that efficiently prevents outages caused by beam misalignment and dynamic blockage while maintaining a low pilot overhead.
Simulation results show that the proposed beamforming scheme improves average rate performance and reduces the amount of outages caused by the brittle THz misalignment process and the particularly severe path loss in the THz band. Moreover, the proposed event-triggered THz channel estimation approach enables connectivity with minimal overhead and reliable communication at THz bands.
\end{abstract}


\section{Introduction}
  A fundamental characteristic of next-generation wireless 6G networks is the migration towards higher frequency bands, namely the terahertz (THz) band (0.1--10 THz)\footnote{The frequency range 100 -- 300 GHz is typically referred to as the sub-THz band, while the unique properties of the THz band are observed above 275 GHz. However, in this work the term THz is used to refer to the overall range 0.1 -- 10 THz.}. Wireless communication links at the THz frequency bands benefit from an abundant bandwidth which enables extremely high data rates (in the order of Tbps) that are essential for future 6G services like extended reality (XR) \cite{chaccour2020ruin} or digital twins \cite{khan2022digital}. However, unleashing the true potential of THz frequency bands necessitates overcoming key THz challenges, stemming from the channel's uncertainty. Particularly, two major factors that restrain the communication at THz bands are the high path loss and the molecular absorption effect \cite{saad2021sevenTHz, akyildiz2018combating}. To compensate the effect of these phenomena, a very narrow beam (so-called pencil beam) is needed to focus the power towards the receiver \cite{sarieddeen2021overview}. However, such a narrow beam makes the communication prone to blockages and beam misalignment and, consequently, it jeopardizes the communication reliability. Indeed, even minute changes of the target direction (in the order of a few degrees or less) can cause communication outages, which becomes particularly prevalent in dynamic use-cases. 
  While this phenomenon can affect mmWave communication, it becomes substantially more pronounced in (sub-)THz systems. Due to the extremely high path loss, the THz band is more suitable for indoor environments with rather short communication distances. That said, in such scenarios, the micro-mobility of users may lead to changes in the angle of departure (AoD) directly affecting the THz pencil beams and their alignment \cite{petrov2020capacity, kokkoniemi2020impact}.  Henceforth, investigating the tradeoff between the pathloss compensation and the mitigation of beam misalignment is substantially crucial for the deployment of THz bands \cite{chen2021mobility}. Indeed, the optimal tradeoff adjustment could ultimately lead to the delivery of \emph{reliable and robust THz links in dynamic environments}, a fundamental necessity for 6G services like XR \cite{chaccour2020ruin}.
  
 In order to maintain system reliability, beam tracking algorithms must provide very precise channel state information (CSI), which comes at the expense of a considerable amount of transmission overhead. This challenge is further exacerbated by the large antenna arrays needed at THz frequencies to form narrow beams. Thus, designing a robust beamforming scheme that can flexibly adjust to CSI uncertainties is a pillar for the guarantee of reliable low-overhead THz communication links.
 Consequently, the tradeoff between providing sufficient communication range with a highly focused beam versus increasing the probability of coverage by generating a wider beam is a key challenge in THz beamforming \cite{chen2021mobility}.
Furthermore, while timely channel estimation is necessary to prevent communication outages, frequent pilot transmissions could induce a significant overhead that restricts the transfer of large amounts of data with low latencies. Hence, pilot based channel measurements should be employed in an efficient manner to handle the highly varying THz channel without violating restrictions on a tolerable overhead amount.
 \subsection{Prior Art}
 The challenge of handling beam misalignment induced by user mobility in THz systems has been addressed in recent articles. For instance, in \cite{chen2021mobility, kokkoniemi2020impact, petrov2020capacity}, the impact of beam misalignment on THz communication performance is investigated for different mobility scenarios. 
These works demonstrate the susceptibility of THz systems to small-scale user mobility and, hence, the need for reliable THz communication schemes with regard to time-varying channels. Beam alignment and tracking approaches have been proposed to improve channel estimation accuracy in mmWave and THz communications (e.g., \cite{attaoui2022initial} and references therein). Since THz communication relies on pencil beams to overcome the path loss, reducing the overhead required for beam training is particularly challenging at THz frequencies.

The authors in \cite{yang2019BTdyntrainfreq} propose a hierarchical codebook-based beam training scheme for unmanned aerial vehicles (UAVs) with a dynamic training frequency, thereby reducing the training overhead. In addition, they study the relationship between beamwidth and the UAV's mobility pattern. A multi-resolution hierarchical codebook is also utilized in \cite{stratidakis2020hierarchical, park2022fast, ning2021unified} to enable low-overhead beam training in mmWave and THz systems. A low-overhead beam tracking algorithm with adaptive tracking rate for mmWave systems is proposed in \cite{liu2020robust}. The authors in \cite{moltchanov2021ergodic} analyze outage probability and spectral efficiency for different beam searching methods in THz systems with micro-mobility. 
 Although methods considered in \cite{yang2019BTdyntrainfreq, stratidakis2020hierarchical, park2022fast, ning2021unified, liu2020robust, moltchanov2021ergodic} are able to reduce the training overhead, the time steps during which beam training is initiated are either periodic or follow a heuristic on-demand or (fixed) threshold-based approach. In contrast, a proactive and reactive event-based strategy as proposed in this paper has the potential to improve balance of communication reliability and overhead efficiency.
Furthermore, while beam tracking approaches can certainly improve channel estimation accuracy and thereby reduce the occurrence of antenna misalignment with a moderate pilot overhead, the beamforming concept itself needs to be robust in the face of remaining channel uncertainties. 

The works in \cite{yang2019BTdyntrainfreq, stratidakis2020hierarchical, park2022fast, ning2021unified} use different beamwidths in the course of hierarchical grid-based search. However, the transmission beamwidth is not adapted to the estimation uncertainty in the time intervals between training.
In \cite{joshi2019tactile}, the beamwidth tradeoff has been studied for mmWave systems, yet instead of optimizing the beamwidth, the beam is widened step by step until a minimum average signal strength is obtained at the receiver.
The relationship between optimal beamwidth and channel uncertainty in a mmWave system is studied in \cite{peng2017robustWB} and \cite{chung2020adaptiveBT}. The authors in \cite{peng2017robustWB} propose a chirp-sequence-based precoder to adapt the beamwidth to the current uncertainty of the user's direction to maximize the ergodic data rate. They show that a wider beam results in a gain of the expected rate if the estimation of the angle of departure (AoD) becomes inaccurate and the signal-to-noise-ratio (SNR) is sufficiently high. While they only consider long-term average rate performance, in a block fading channel, however, transmission failures can occur as a consequence of beam misalignment, which severely impede continuous data transfer. Hence, communication reliability should be captured by the performance metrics, e.g., by considering outage probability along with the rate.
The idea of increasing robustness by dynamically adapting the beamwidth to the current channel uncertainty has been applied to a mmWave beam tracking scenario in \cite{chung2020adaptiveBT}. Therein, the beamwidth is adjusted by activating only part of the antenna array, while the selection of the number of active antennas follows a heuristic approach based on the angular deviation. 
However, the work in \cite{chung2020adaptiveBT} is not suitable for THz systems, as it does not include the adaptation of the beamwidth to the channel gain, which is highly affected by user mobility as well as the molecular absorption effect.

In summary, while there has been works on related ideas in mmWave bands \cite{joshi2019tactile, peng2017robustWB, chung2020adaptiveBT}, these existing approaches are not effective for simultaneously providing high data rates and high reliability in THz systems. This is because the tradeoff between path loss compensation and beam alignment arising from the peculiarities of THz bands is not resolved with respect to data rate and outage objectives. 
Moreover, the existing works on low-overhead tracking propose heuristic approaches rather than optimizing CSI estimation time intervals. Thus, these schemes do not ensure compliance of overhead limitations while simultaneously adjusting to the channel, mobility pattern, and the transmission scheme.

 \subsection{Contributions}
The main contribution of this paper is the design of a new, low-overhead beamforming and tracking scheme that enhances communication reliability while addressing the peculiar challenges of the THz band. While mmWave networks often rely on codebook-based beam search, this approach is not suitable for THz systems. In contrast to mmWave bands, THz bands suffer from a considerably more severe path loss, which requires narrow beams, i.e., a high-resolution codebook. Searching over such a large number of beams, however, induces a lot of overhead \cite{attaoui2022initial}. 
Thus, we propose a beamforming design that accounts for the highly varying nature of the THz channel by addressing the tradeoff between THz path loss compensation and beam alignment, and yields an operation with minimal overhead, yet robust and reliable communication (see Fig. \ref{fig:motivation}).
In this work, we consider downlink communications in an indoor wireless THz network with a dense base station (BS) deployment and multiple mobile user equipments (UEs) that are subject to dynamic channel blockage as shown in Fig. \ref{fig:sys_model}:

\begin{itemize}
        \item \emph{Multi-objective optimization:}
        We aim at optimizing the beamformer and the time steps at which pilot-based channel estimation is performed, in order to provide reliable communication at high data rates, while maintaining a low overhead (see Fig. \ref{fig:motivation}). We formulate an optimization problem that aims at jointly maximizing the expected data rate and minimizing the outage probability, subject to constraints on the transmit power and the long-term average overhead. Applying linear scalarization to the multi-objective problem enables balancing high rate and reliability requirements according to the application. The problem is then split into two subproblems to solve for the beamformer and the pilot event times separately.
        
        \begin{figure}[]
    \centering
    \subfigure[Rate-reliability tradeoff illustration]{\label{fig:motivation}
   \includegraphics[]{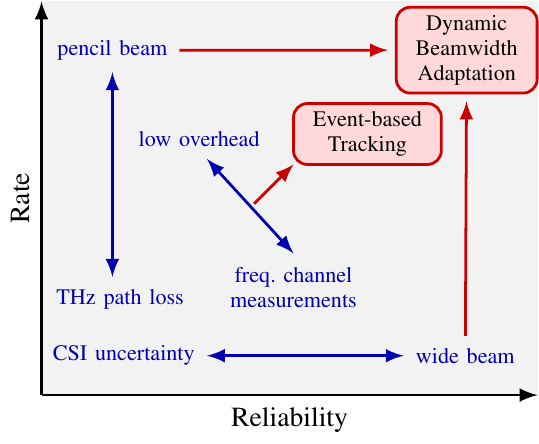}}
   \hspace{1cm}
    \subfigure[Indoor THz communication network]{\label{fig:sys_model}
    \includegraphics[]{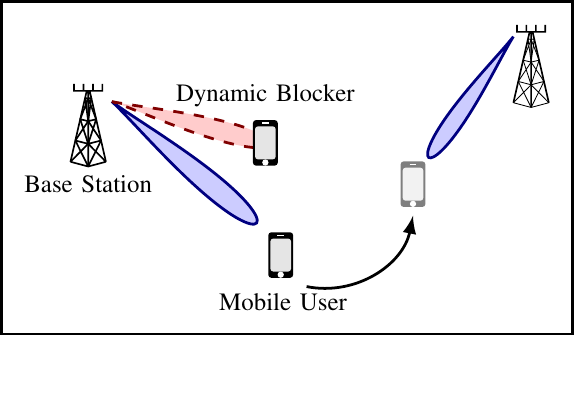}}
    \caption{Considered problem illustration: \subref{fig:motivation} Impact of beamwidth and overhead with regard to the rate-reliability tradeoff in THz systems. Our proposed approach comprising a reliable beamforming scheme and an optimized tracking via event times enable the counteraction of path loss and beam misalignment, \subref{fig:sys_model} Downlink transmission to a mobile UE. Other users are potential blockers to the considered communication link, while other BSs can take over serving the user if needed, e.g., if a blockage occurs.\vspace{-.3cm}}
    \label{fig:Problem}
\end{figure}

        \item \emph{Path loss vs. misalignment tradeoff:}
        We propose a novel beamforming scheme that is reliable in front of CSI uncertainties. In order to reduce the computational complexity of our precoding, we propose a parameterized beamformer with adjustable beamwidth. Combining this with a small-angle approximation allows us to solve the optimization problem in advance and generate a lookup table for the optimal beam parameters, which dynamically adjust the beamwidth depending on the current channel uncertainty and path loss. 
        
        \item \emph{Event-trigger for overhead reduction:}
        To ensure reliable communication by further minimizing beam misalignments without violating the average overhead constraint, we adopt the concept of event-triggered communication, that enables more efficient scheduling of pilots and control-related data (e.g., in \cite{karacora2020energy}). Instead of periodically transmitting pilot signals to estimate the channel, the interval between consecutive channel measurements is dynamically adapted to the current system state.
        Here, we adopt a Lyapunov optimization framework to determine the time steps, at which the BS receives updates on the UE's current direction. This enables a flexible system capable of not only reacting to outage events when needed but preventing outages by ensuring sufficiently accurate CSI while still complying with a given average overhead constraint.
        
        \item \emph{Beamforming scheme analysis:}
        We analyze the proposed beamforming scheme by numerically calculating the Pareto boundary of the two objectives, namely expected data rate and outage probability, for a general beamformer and compare it to the achievable Pareto region of our proposed parameterization. We further gain insights regarding the optimal beamwidths for different BS-UE distances and uncertainty of the UE's position as well as the impact of the molecular absorption effect in the THz band. Here, we observe that the optimal beamforming strategy differs significantly for the two considered objectives. Moreover, our approach is shown to outperform state-of-the-art variable beamwidth schemes.
        
        \item \emph{Analysis of event-based tracking procedure:}
        The performance of our event-based tracking scheme combined with the proposed beamforming approach is evaluated and compared to a non-robust periodic scheme. Our approach is shown to significantly reduce the amount of outage events while requiring much less pilot overhead. 
\end{itemize}

The rest of this paper is organized as follows. In Section~\ref{sec:systemmodel} the channel model and UE's mobility model are introduced. Then, in Section~\ref{sec:beamformer_opt}, the proposed robust beamforming scheme is presented. Section~\ref{sec:event+tracking} describes the event-based channel estimation and tracking approach. Section~\ref{sec:numerical} presents the simulation results and, finally, conclusions are drawn in Section~\ref{sec:conc}.

\textbf{Notation:} 
Vectors and matrices are denoted by boldface lowercase and uppercase letters, respectively. The operators $\mathbb{E}[\cdot]$, $|\cdot|$ and $\lfloor \cdot \rfloor$ represent the expectation, the absolute value and the floor function, respectively. $\vec{X}^T$ and $\vec{X}^H$ are the transpose and the hermitian, while $[\mat{X}]_{m,n}$ denotes the element in the $m$-th row and $n$-th column of $\mat{X}$.

\section{System Model}\label{sec:systemmodel}
In this work, we consider a multiple-input-multiple-output (MIMO) THz communication system, where multiple BSs, which are densely deployed in an indoor area, are transmitting data to multiple mobile UEs.
Each BS (UE) is equipped with a uniform linear array (ULA) consisting of $N_\mathrm{t}$ ($N_\mathrm{r}$) antennas.
We focus on the downlink of a single mobile UE $i$ and its associated serving BS $j$ (see Fig. \ref{fig:sys_model}). From this perspective, other users take the role of potential blockers. Hence, communication is jeopardized for three main reasons, namely beam misalignment caused by the UE's changing position, dynamic blockage induced by other users, and the user moving out of the communication range of the BS. To cope with these impairments, the BS-UE channel is estimated based on pilot measurements on a regular basis. Additionally, the mobile user can be handed over to another BS if coverage is disrupted. 

\subsection{Channel Model}

We adopt the Saleh-Valenzuela channel model, that is widely used for THz communications (e.g., see \cite{stratidakis2020hierarchical, ning2021unified, sarieddeen2021overview}) and generally consists of one line-of-sight (LOS) path and a few reflection paths.
However, since at the THz band the attenuation induced by scattering is more than 20 dB compared to the LOS path \cite{boulogeorgos2018performance}, we neglect the non-line-of-sight (NLOS) component and only consider the dominant LOS path in our channel model as follows:
\begin{equation*}
    \mat{H}_k = \gamma_k \eta(d_k) \vec{a}_\mathrm{r}(\varphi_{\mathrm{r},k}) \vec{a}_\mathrm{t}^H(\varphi_{\mathrm{t},k}),
\end{equation*}
where $k$ is the time index, while $\eta(\cdot)$, $\varphi_\mathrm{t}$ and $\varphi_\mathrm{r}$ represent the path gain, the AoD and angle of arrival (AoA) of the LOS path, respectively.
The distance between the UE and its associated BS is denoted by $d_k$, while $\vec{a}_\mathrm{t}(\cdot)$ and $\vec{a}_\mathrm{r}(\cdot)$ are the transmit and receive array response vectors, respectively. We capture the effect of blockage by defining a random binary variable $\gamma_k$, that is equal to zero if the LOS path between the considered BS-UE pair is blocked at time step $k$ and equal to one otherwise. Note that the blockage model is explained in more detail in Section \ref{sec:blockage}. 
Without loss of generality, we assume the antenna spacing to be half of a wavelength. Hence, the array response vectors are defined as:
\begin{equation}\label{steering_vec}
    \vec{a} (\varphi) = \frac{1}{\sqrt{N}}\left[1,~ e^{j \pi \sin(\varphi)},\dots, e^{j \pi (N-1) \sin(\varphi)}\right]^T.
\end{equation}
At THz frequencies, in addition to the free space propagation loss, the path loss is highly affected by molecular absorption and re-radiation. Hence, the total channel gain is given by \cite{chaccour2020HRLLC}:
\begin{equation}
	\eta(d_k) = \frac{c}{4\pi f d_k} e^{-\frac{1}{2}K(f) d_k},
\end{equation}
where $f$ is the carrier frequency, $c$ is the speed of light, and $K(f)$ represents the overall absorption coefficient of the medium. We obtain $K(f)$ for the frequency range of 100 -- 450 GHz based on the model presented in \cite{kokkoniemi2021line}.
At time step $k$, the BS transmits the symbol $s_k$ with $\mathbb{E}[|s_k|^2] = 1$. The received signal will be given by:
\begin{equation*}
    y_k = \vec{w}^H_k \mat{H}_k \vec{f}_k s_k + n_k,
\end{equation*}
where the precoder and combiner are denoted by $\vec{f}_k$ and $\vec{w}_k$, respectively, and $n_k \sim \mathcal{CN}(0, \sigma_\mathrm{n}^2)$ is additive white Gaussian noise (AWGN) where $\sigma_\mathrm{n}^2 = N_0 + P_\mathrm{max} \left(\frac{c}{4\pi f d_k}\right)^2(1-e^{-K(f)d_k})$ is the sum of the thermal noise power $N_0$ and molecular absorption noise caused by molecular re-radiation \cite{chaccour2020HRLLC, kokkoniemi2016discussion}.
The achievable data rate with bandwidth $W$ can be written as:
\begin{equation}\label{rate_general}
    R_k = W \log_2\left(1 + \frac{1}{\sigma_\mathrm{n}^2} \vec{w}_k^H \mat{H}_k \vec{f}_k \vec{f}_k^H \mat{H}_k^H \vec{w}_k\right).
\end{equation}
Note that in \eqref{rate_general}, the effect of interference caused by other devices is neglected given the generally narrow beams utilized in THz frequency bands.

\subsection{Blockage model} \label{sec:blockage}
Given that the THz frequency band suffers from high reflection losses, the communication relies mainly on the LOS component, which also suffers from high penetration losses. As a result, the LOS links are highly susceptible to blockages, which jeopardizes the system reliability. 
There are two different types of blockages: \emph{static} blockage, caused by, e.g., room architecture and furnishing, and \emph{dynamic} blockage, which results from other users temporarily blocking the LOS link of the tagged BS with their body while moving around the room.
While we assume that static blockages can easily be avoided by appropriate positioning of the BSs in an indoor scenario, we still have to deal with dynamic blockages induced by user mobility in a dense network. We model this effect as an M/M/$\infty$ queuing system \cite{chaccour2020HRLLC, jain2019block}. More specifically, the occurrence of dynamic blockages is modeled as a Poisson process with an arrival rate $\kappa_\mathrm{B}$ blockers/sec and an exponentially distributed blockage duration with parameter $\mu_\mathrm{B}$. That is, the binary blockage variable $\gamma_k$ follows an exponential on-off process with $\kappa_\mathrm{B}$ and $\mu_\mathrm{B}$ as the blocking and unblocking rate, respectively. 
The corresponding blocking and unblocking probabilities are 
\begin{equation} \label{blockage_prob}
    P(\gamma_k=0|\gamma_{k-1}=1) = 1 - e^{-\kappa_\mathrm{B}(d_k) T_\mathrm{s}}~~ \text{and}~~  P(\gamma_k=1|\gamma_{k-1}=0) = 1 - e^{-\mu_\mathrm{B} T_\mathrm{s}},
\end{equation}
where $T_\mathrm{s}$ is the length of a single time slot.
While the unblocking rate $\mu_\mathrm{B}$ is assumed to be a constant parameter known by the BS, the blockage arrival rate depends on the distance between BS and UE and is obtained by:
\begin{equation*}
    \kappa_\mathrm{B}(d_k) = \frac{2}{\pi} \lambda_\mathrm{B} v_\mathrm{B} \frac{h_\mathrm{B} - h_\mathrm{UE}}{h_\mathrm{BS} - h_\mathrm{UE}} d_k,
\end{equation*}
where $\lambda_\mathrm{B}$ denotes the density of dynamic blockers per m$^2$ and $v_\mathrm{B}$ is the blockers velocity. Here, $h_\mathrm{B}$, $h_\mathrm{UE}$, and $h_\mathrm{BS}$ represent, respectively, the height of the blocker, the considered UE, and the BS. The mobility of the users is modeled next. 

\subsection{Mobility Model}
We model the UE's mobility as a random walk \cite{petrov2020capacity, chaccour2020risk}, where the steps in $x$- and $y$-direction are independently and identically distributed (i.i.d.) as $\mathcal{N}(0, \sigma_\mathrm{m}^2)$. For simplicity, we consider the trajectory of the UE in the horizontal $x$-$y$-plane only, while omitting the height. Under these assumptions the $x$- and $y$-component of the UE's location after $M$ steps (i.e. the sum of $M$ i.i.d. zero-mean Gaussian steps) follows a normal distribution with zero mean and variance $M \sigma_\mathrm{m}^2$ \cite{sanguinetti2014large}. Assuming that the users in an indoor scenario often times do not walk towards a specific destination and frequently change directions, the random walk scheme arises as a useful mobility model for our considered scenario. Moreover, beyond the need to know the average step size, specific knowledge of the UE's movement behaviour is not required at the BS. 

Given that the UE's mobility affects the direction of the LOS path, the BS relies on regular channel estimations at the cost of pilot overhead to adjust the beam accordingly. In order to capture the intermittent CSI updates, we define a binary variable $q_k$, which is equal to one whenever the BS obtains a new channel estimate and is equal to zero in between those updates. 
As the channel depends directly on the AoD, the BS is assumed to obtain perfect knowledge of the current AoD and distance of the UE when $q_k = 1$.

Let $\vec{p}_k = [d_k \cos(\varphi_{\mathrm{t},k}),~d_k \sin(\varphi_{\mathrm{t},k})]^T$ be the position vector of the UE at time step $k$, where $d_k$ and $\varphi_{\mathrm{t},k}$ denote the current UE's distance to the BS and the AoD, respectively. Then, the position estimate at the BS will be given by:
\begin{equation*}
    \vec{\hat{p}}_k = \begin{cases}\vec{p}_k, & \text{if}~ q_k=1,\\ \vec{\hat{p}}_{k-1}, & \text{if}~ q_k = 0.\end{cases}
\end{equation*} 
Given that the user mobility is modeled as a random walk with Gaussian step size, the estimation error $\vec{p}_k - \hat{\vec{p}}_k$ is also Gaussian distributed with zero mean and covariance matrix $\boldsymbol{\Sigma}_{\mathrm{p},k} = \sigma_{\mathrm{p},k}^2 \mat{I}$. We assume that the BS is able to obtain the step size variance $\sigma_\mathrm{m}^2$ of the user to find the variance $\sigma_{\mathrm{p},k}^2 = M_k \sigma_\mathrm{m}^2$, where $M_k$ is the number of time steps since the BS received the most recent update of the user's position.

\begin{figure}
    \centering
    \includegraphics[]{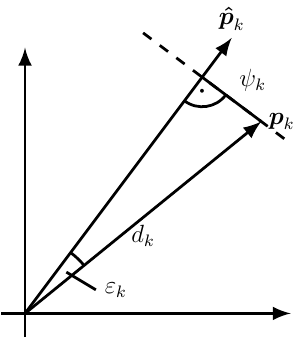}
    \caption{Geometry of expected user position vector $\vec{\hat{p}}_k$, actual position $\vec{p}_k$ and AoD estimation error $\varepsilon_k$, assuming that the BS is located at the origin.}
    \label{fig:geometry}
\end{figure}
Let $\varepsilon_k = \varphi_{\mathrm{t},k} - \hat{\varphi}_{\mathrm{t},k}$ be the AoD estimation error. 
As the distribution of $\varepsilon_k$ is quite complex (see \cite{aalo2007phasedistr} for the exact distribution), it can be approximated by a normal distribution for small $\varepsilon_k$ as follows\footnote{Note that the assumption of small AoD errors is reasonable, since even small deviations of the AoD are critical given the narrow beams in THz communication.}: From Figure \ref{fig:geometry}, we obtain:
\begin{equation*}
    \frac{\psi_k}{d_k} = \sin(\varepsilon_k) \approx \varepsilon_k,
\end{equation*}
where  $\psi_k$ is the component of the location error $\vec{p}_k - \vec{\hat{p}}_k$, which is orthogonal to $\vec{\hat{p}}_k$ and $d_k \gg \psi_k$. Since the distribution of the position estimation error is circular symmetric, we know that $\psi_k \sim \mathcal{N}(0,\sigma_{\mathrm{p},k}^2)$. By further approximating $d_k$ by $\hat{d}_k$, for small $\varepsilon_k$, we can assume $\varepsilon_k \sim \mathcal{N}(0, \sigma_{\mathrm{p},k}^2/\hat{d}_k^2)$. Note that the AoD estimation error following a normal distribution is a common assumption in other works on beam tracking as well, such as in \cite{heath2016beamEKF, chung2020adaptiveBT, jayaprakasam2017BT_EKF, ge2019unscented}. Hence, with $\sigma_{\varepsilon,k}^2 = \sigma_{\mathrm{p},k}^2/\hat{d}_k^2$, we define the probability density function (PDF) of $\varepsilon_k$ as:
\begin{equation} \label{pdf_eps}
    g_k(\varepsilon_k) = \frac{1}{\sqrt{2 \pi \sigma_{\mathrm{\varepsilon},k}^2}} \exp\left(- \frac{\varepsilon_k^2}{2 \sigma_{\varepsilon,k}^2} \right).
\end{equation}

\subsection{Tracking and Problem Statement}
Our goal is to design a beamforming scheme that enables reliable communication despite the uncertainty of the user's location while maintaining a low channel estimation overhead. The problem is formulated from the perspective of a single UE and its associated BS, while other users are considered as potential dynamic blockers and other BSs might take over serving the UE in case of link blockage or if the user moves out of the BS's communication range. 
The BS obtains information of the current CSI and UE's location at intermittent time steps through pilot signal measurements. These pilot transmissions are initiated in a non-periodic event-triggered manner to comply with a maximum average overhead constraint. In between the CSI updates, the BS transmits data to the UE while adjusting the beamformer in every time slot based on the available statistical CSI.
Note that \emph{communication outages} caused by insufficient signal strength at the receiver can be induced in multiple ways: 
First, CSI at the BS is not always sufficiently accurate, since user mobility in a non-static environment leads to a fast varying channel. As a result, beams are not perfectly aligned, which can cause communication outages. This effect is aggravated due to the very narrow beams commonly used to overcome the severe path loss in THz channels. Second, the high penetration loss at THz bands leads to blockages, particularly caused by other users in a mobile environment. Beyond that, an outage can occur as a result of an excessive path loss when the UE moves out of the THz communication range of the BS.
While the BS can react to outages with a new channel measurement or a handover, we aim at designing a robust beamforming scheme, that provides high data rates despite inaccurate CSI and severe THz path loss and reduces the probability of outages caused by beam misalignment. We define an outage as the event of the data rate $R_k$ falling below a given target rate $R_\mathrm{min}$. 

A suitable performance metric to consider in communication scenarios with imperfect and outdated CSI at the transmitter is the expected data rate. By maximizing $\mathbb{E}[R]$, a relatively high system throughput can be provided on average despite channel uncertainties. However, when solely considering the expected rate, interruptions in communication in the form of outages or temporary rate decline can be concealed by this metric. Indeed, beyond the need for high data rates, many THz communication use-cases also depend on a continuous and timely data transfer. For instance, XR applications require high rates to deliver the visual content to the users, but for a seamless user experience the content packet transmissions must be highly reliable as well. A useful metric to capture instantaneous violations of these QoS demands is the probability of outage. A communication scheme designed to reduce the outage probability ensures consistent data transfer at the expense of total throughput. Due to its threshold-based definition, outage probability is also vulnerable to channel uncertainties. 
Given the highly susceptible THz channel and the user mobility in our considered communication setup, outages or violation of QoS requirements are likely to occur. In order to provide high data rates despite channel uncertainties, yet avoid outages caused by beam misalignment, we formulate a multi-objective optimization problem \cite{bjornson2014multiobjective, xu2014ergodic}.
In addition to the beamforming scheme, we also aim at optimizing the time steps, at which a new pilot measurement is performed, and thereby making the tracking procedure more efficient. This allows for timely CSI updates to prevent outages, while a low overhead can be maintained. 
Hence, our goal is to optimize the beamformer and the pilot measurement times with regard to the expected data rate and the outage probability, subject to constraints on the transmit power and the long-term average pilot overhead. Mathematically, we have the following multi-objective optimization problem:
\begin{align} \label{opt_MO}
        \max_{\vec{f}_k, ~q_k} \quad &\left(\mathbb{E} \left[R_k|\vec{f}_k,q_k \right] ,~  -\mathrm{Pr}(R_k < R_\mathrm{min}|\vec{f}_k,q_k) \right)\\
        \text{s.t.}\quad &\vec{f}_k^H\vec{f}_k \leq P_\mathrm{max},\label{power_const}\\
        & \lim_{K\rightarrow \infty} \frac{1}{K}\sum_{k=1}^{K} q_k \leq r_\mathrm{q}\label{overhead_const},
\end{align}
where $P_\mathrm{max}$ denotes the maximum transmit power and $r_\mathrm{q}$ represents the allowable average channel estimation overhead.
Note that in the evaluation, we demonstrate that each of the two objectives considered by itself leads to substantially different beamforming strategies.
The problem is converted into a single-objective optimization problem through scalarization \cite{bjornson2014multiobjective} as follows:
\begin{equation} \label{opt_SO}
    \begin{aligned}
        \max_{\vec{f}_k,~q_k} \quad & g_\alpha(\vec{f}_k,q_k) \\
        \text{s.t.}\quad & \eqref{power_const},~\eqref{overhead_const},
    \end{aligned}
\end{equation}
in which the objective function is chosen to be a weighted sum of the two objectives, given by:
\begin{equation} \label{objective}
    g_\alpha(\vec{f}_k,q_k) = \alpha~\frac{\mathbb{E} \left[R_k|\vec{f}_k,q_k \right]}{R_\mathrm{max}} - (1-\alpha)  \mathrm{Pr}(R_k < R_\mathrm{min}|\vec{f}_k,q_k),
\end{equation}
where $\alpha \in [0,1]$ is a design parameter, which balances the two objectives.
In order to make the two metrics comparable, in \eqref{objective}, the expected data rate is normalized by $R_\mathrm{max}$, which is an approximation of the maximum expected rate \cite{xu2014ergodic}. This makes both objectives in the weighted sum dimensionless values between 0 and 1. $R_\mathrm{max}$ is chosen to be the achievable rate with perfect CSI available at the transmitter. Hence, it is obtained as follows:
\begin{equation} \label{R_max}
    R_\mathrm{max} = W \log_2 \left(1 + \frac{P_\mathrm{max} \eta^2(\hat{d}_k)}{\sigma_\mathrm{n}^2}\right).
\end{equation}

In the following two sections, the optimization problem in \eqref{opt_SO} is split into two subproblems, optimizing the beamforming vector $\vec{f}_k$ first, and then deriving a dynamic condition that triggers the pilot measurements by optimizing the channel estimation variable $q_k$.

\section{Reliable Variable-Beamwidth Precoding} \label{sec:beamformer_opt}
Given that the BS's location is fixed, we assume that the UE is aware of its own position relative to the BS, i.e., the current AoA $\varphi_{\mathrm{r},k}$ is available at the UE. Thus, the UE will apply maximum ratio combining to the received signal, so that:
\begin{equation}\label{w_MRC}
\vec{w}_k = \frac{\vec{a}_{\mathrm{r},k}(\varphi_{\mathrm{r},k})}{||\vec{a}_{\mathrm{r},k}(\varphi_{\mathrm{r},k})||}.
\end{equation}

 Note that when $q_k=1$, i.e., a new channel measurement has been performed at the beginning of time step $k$ and, hence, the BS is assumed to have perfect CSI, a maximum ratio transmitting strategy would be optimal. Therefore, in the following, we solve the precoding optimization problem for the case of imperfect CSI, i.e., under the assumption $q_k = 0$ with the overhead constraint \eqref{overhead_const} becoming irrelevant for this subproblem. Similarly, since the rate is always zero when a blockage occurs, i.e., $\gamma_k=0$, blockages can be ignored during the beamforming optimization. The optimization of $q_k$ including blockages is covered in Section \ref{sec:event}. 
 Thus, with these assumptions and the combiner in \eqref{w_MRC}, the subproblem for beamforming optimization can be written as:
\begin{equation} \label{opt_BF}
    \begin{aligned}
        \max_{\vec{f}_k} \quad & g_\alpha(\vec{f}_k,q_k=0) \\
        \text{s.t.}\quad & \eqref{power_const},
    \end{aligned}
\end{equation}
while the rate expression in \eqref{rate_general} reduces to
\begin{equation}\label{rate1}
    R_k = W \log_2\left(1 + \frac{\eta^2(d_k)}{\sigma_\mathrm{n}^2} \left\lvert\vec{a}_\mathrm{t}^H(\varphi_{\mathrm{t},k})\vec{f}_k\right \rvert^2\right).
\end{equation}

Note that \eqref{opt_BF} is a challenging problem, since it is non-convex and the expectation operator in the objective function cannot be solved in closed form. Additionally, the optimal precoder highly depends on the communication distance and the AoD distribution, which requires constant recalculation while the UE is moving. Therefore, in what follows, we propose a parameterized precoder based on two real-valued scalar parameters to control the width and shape of the beam, respectively. Besides reducing complexity and improving scalability of the optimization, this allows us to precalculate a look-up table for the beam parameters, which can be used during communication. The parameterization is based on the idea, that the tradeoff between increasing communication range on the one hand and improving robustness towards AoD uncertainty on the other hand can be tackled by dynamic beamwidth adaptation. 

 \begin{figure}[tb]
     \centering
     \includegraphics[]{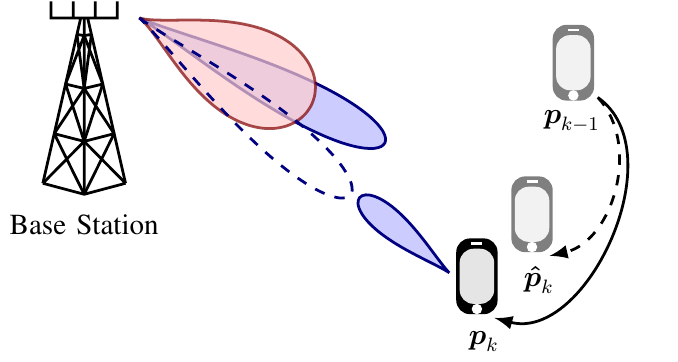}
     \caption{Illustration of beam misalignment induced by user mobility.}
     \label{fig:beam_misalignment}
 \end{figure}
 
Figure \ref{fig:beam_misalignment} illustrates the potential benefit from variable-beamwidth precoding in the presence of user mobility. As the UE moves from position $\vec{p}_{k-1}$ to $\vec{p}_k$, the BS's estimate of the UE's position at time step $k$ is $\vec{\hat{p}}_k$. Forming a narrow beam (shown in blue) toward the expected user direction would most likely lead to an outage as the transmit and receive beams are misaligned and very low power is recieved by the UE. However, when forming a wider beam (shown in red), the UE's actual direction would still be covered by the main lobe, making communication more reliable despite AoD uncertainty. 
In particular, we expect a wider beam to be more robust if the AoD estimation error variance is high, while a narrow beam should be preferred when the AoD estimation is sufficiently accurate. 
However, since the THz band suffers from particularly severe path loss depending on the distance $d_k$ and the molecular absorption coefficient, ensuring sufficiently high signal strength at the receiver is also a key factor in the beamformer design. Hence, we are facing a tradeoff between increasing the probability of covering the user and enhancing the received signal strength.
We tackle this challenge by considering the optimization problem \eqref{opt_BF} with a parameterized variable beamwidth precoder, which we propose next.

\begin{remark} \label{remark_general_sol}
    As previously explained, we propose a parameterization of the precoder for computational complexity and scalability reasons. However, to justify the accuracy of our parameterization, we also solve \eqref{opt_BF} for a general beamforming vector $\vec{f}_k$, and compare the achievable Pareto region of the general and the parameterized solution in the simulation section. To find a local optimum of the general problem, we apply gradient ascent method. However, since \eqref{opt_BF} is a non-convex problem, the numerical optimization leads to different local optima depending on the initialization of $\vec{f}_k$. Therefore, we repeat the gradient ascent for different initial values and pick the best locally optimal solution. Although we cannot claim our general solution to be a global optimum, we gain insights regarding the achievability region and validate the usefulness of our parameterization.
\end{remark}

\subsection{Adaptive Beamwidth Precoder}\label{sec:sinc_BF}

In this section, we derive a dynamic beamwidth adaptation scheme based on a parameterized precoder. Notice that a wide beam can be formed by adding up multiple beams, which are slightly offset from the expected UE's direction, as it has been done for a multi-resolution codebook design in \cite{love2017multires}. In contrast to \cite{love2017multires}, we sum up infinitely many beams within a certain angular range parameterized by $v \in [0,1]$, which leads to a precoder of the following form:

\begin{equation}
    \vec{f}(v,\omega) = \frac{\beta}{2v} \int_{-v}^v \vec{u}(\hat{\varphi},\xi) e^{j\omega \xi} \mathrm{d}\xi,
\end{equation}
where $\beta$ is a scaling factor that ensures the transmit power constraint and \begin{equation}
    \vec{u}(\hat{\varphi},\xi) = \left[1,~ e^{j\pi \left(\sin(\hat{\varphi}) - \xi\right)} ,~\dots,~e^{j\pi (N-1)\left(\sin(\hat{\varphi}) - \xi\right)}\right]^T.
\end{equation}
The additional phase shift given by the parameter $\omega$ helps optimize the beam shape.
Hence, the $n$-th component of the precoding vector can be determined in closed form, as follows: 
\begin{equation}\label{f_sinc}
    \begin{split}
        [\vec{f} (v,\omega)]_n &= \frac{\beta}{2v} \int_{-v}^v e^{j\pi n\left(\sin(\hat{\varphi}) - \xi\right)} e^{j\omega \xi} \mathrm{d}\xi\\
        &= \frac{\beta}{2v}  \frac{e^{j(\omega - \pi n)v} - e^{-j(\omega-\pi n)v}}{j(\omega-\pi n)} e^{j\pi n \sin(\hat{\varphi})}\\
        &= {\beta} \frac{\sin\left((\omega- \pi n)v\right)}{(\omega-\pi n)v} e^{j\pi n \sin(\hat{\varphi})}.
    \end{split}
\end{equation}
Note that this beamforming structure is related to the Slepian sequence used for bandpass filter design, where the energy within a certain frequency interval is maximized \cite{love2017multires}.

We adopt the precoder \eqref{f_sinc} when considering the optimization problem \eqref{opt_BF}. In particular, when inserting \eqref{f_sinc} and \eqref{steering_vec} in \eqref{rate1}, we obtain the following rate expression:
\begin{equation}\label{rate_vw}
    R_k = W \log_2\left(1 + \frac{\beta_k^2\eta^2({d}_k)}{N_\mathrm{t} \sigma_n^2} \left\lvert \sum_{n=0}^{N_\mathrm{t}} \frac{\sin\left((\omega_k- \pi n)v_k\right)}{(\omega_k-\pi n)v_k} e^{j\pi n\left(\sin(\hat{\varphi}_{\mathrm{t},k}) - \sin(\hat{\varphi}_{\mathrm{t},k} + \varepsilon_k)\right)}\right\rvert^2 \right),
\end{equation}
where $\varepsilon_k=\varphi_{\mathrm{t},k} - \hat{\varphi}_{\mathrm{t},k}$ is the AoD estimation error with PDF given in \eqref{pdf_eps}.
Hence, the optimization problem \eqref{opt_BF} reduces to:
\begin{equation}\label{opt_BF_vw}
    \begin{aligned}
        \max_{v_k,\omega_k, \beta_k}\quad &  \alpha~\frac{\mathbb{E} \left[R_k|\vec{f}_k(v_k,\omega_k),q_k=0 \right]}{R_\mathrm{max}} - (1-\alpha)  \mathrm{Pr}(R_k < R_\mathrm{min}|\vec{f}_k(v_k,\omega_k),q_k=0)\\
        \text{s.t.}\quad &\beta_k^2 \sum_{n=0}^{N_\mathrm{t}-1} \frac{\sin^2\left((\omega_k- \pi n)v_k\right)}{(\omega_k-\pi n)^2v_k^2} \leq P_\mathrm{max}.
    \end{aligned}
\end{equation}
Next, we derive the approximate expressions for the two objectives, namely the expected rate and the outage probability, before numerically solving \eqref{opt_BF_vw}.

First, we assume $d_k \approx \hat{d}_k$ and consider only the distribution of the AoD when applying the expectation operator to the data rate in \eqref{rate_vw}. However, the expectation can still not be easily solved in closed form and is therefore calculated numerically using an integral expression. Note that when maximizing the expected data rate, the optimal beam parameters depend on the SNR (including transmission power, path loss and noise power) as well as the distribution of the AoD $\varphi_{\mathrm{t},k}$. In order to enable a low-complexity look-up table based offline calculation of the optimal beamformer, we combine the AoD estimate $\hat{\varphi}_{\mathrm{t},k}$ and the AoD error variance $\sigma_\varepsilon^2$ in a single variable by means of the following approximation:
 Under the assumption that $\varepsilon_k$ is small, and utilizing the approximations $\sin(x) \approx x$ and $\cos(x) \approx 1$ for very small $x$, we have:
\begin{equation}\label{angle_approx}
\begin{split}
    \sin(\hat{\varphi}) - \sin(\hat{\varphi} + \varepsilon) &= \sin(\hat{\varphi}) - \left[\sin(\hat{\varphi})\cos(\varepsilon) + \cos(\hat{\varphi})\sin(\varepsilon)\right] \approx \cos(\hat{\varphi}) \varepsilon.
\end{split}
\end{equation}
Let $\tilde{\varepsilon}_k = \cos(\hat{\varphi}_{\mathrm{t},k})\varepsilon_k$. Then, $\tilde{\varepsilon}_k \sim \mathcal{N}(0, \tilde{\sigma}_{\varepsilon_k}^2)$ with $\tilde{\sigma}_{\varepsilon_k}^2 = \cos(\hat{\varphi}_{\mathrm{t},k})^2 \sigma_{\varepsilon_k}^2$.
Hence, the expected rate can be written as:
\begin{equation}\label{E_rate_vw}
    \begin{aligned}
       \mathbb{E}[R_k] &\approx W \int_{-\pi/2}^{\pi/2}\log_2\left(1 + \frac{\beta_k^2\eta^2(\hat{d}_k)}{N_\mathrm{t} \sigma_n^2} \left\lvert\sum_{n=0}^{N_\mathrm{t}-1} \frac{\sin\left((\omega_k- \pi n)v_k\right)}{(\omega_k-\pi n)v_k} e^{j\pi n \tilde{\varepsilon}_k} \right\rvert^2\right) g_k(\tilde{\varepsilon}_k) \mathrm{d}\tilde{\varepsilon}_k.
    \end{aligned}
\end{equation}

 Next, we consider the outage probability. 
Since a closed form expression cannot be easily obtained, we use a logistic function based approximation as objective instead \cite{zhou2020fairness}:
\begin{equation}
    P_{\mathrm{out},k} = \text{Pr}(R_k < R_\mathrm{min}) \approx \mathbb{E} \left[\frac{1}{1 + \exp(-\theta (R_\mathrm{min} - R_k))}\right],
\end{equation}
where $\theta$ is a smoothness parameter to adjust the approximation error.
Hence, with \eqref{rate_vw} and \eqref{angle_approx}, we approximate the outage probability as follows:
\begin{equation} \label{P_out_vw}
    \begin{aligned}
        &\text{Pr}(R_k<R_\mathrm{min}) \approx \mathbb{E} \left[\left(1 + 2^{-\theta R_\mathrm{min}/W}\left({1 + \frac{\eta^2(\hat{d}_k)}{\sigma_\mathrm{n}^2} \left\lvert\vec{a}_\mathrm{t}^H(\varphi_{\mathrm{t},k})\vec{f}(v_k,\omega_k)\right\rvert^2 }\right)^\theta\right)^{-1} \right]  \\
        &\approx \bigintsss_{-\pi/2}^{\pi/2} \left(1 + 2^{-\theta R_\mathrm{min}/W}\left(1 + \frac{\beta_k^2\eta^2(\hat{d}_k)}{N_\mathrm{t} \sigma_n^2} \left\lvert\sum_{n=0}^{N_\mathrm{t}-1} \frac{\sin\left((\omega_k- \pi n)v_k\right)}{(\omega_k-\pi n)v_k} e^{j\pi n \tilde{\varepsilon}_k} \right\rvert^2\right)^\theta\right)^{-1}  g_k(\tilde{\varepsilon}_k) \mathrm{d}\tilde{\varepsilon}_k.
    \end{aligned}
\end{equation}

 Using both \eqref{E_rate_vw} and \eqref{P_out_vw}, we solve Problem \eqref{opt_BF_vw} using a particle swarm optimization method \cite{wang2018particle}.
The optimal beam parameters $v_k$ and $\omega_k$ can be computed for varying $\sigma_{\tilde{\varepsilon}_k}$ and $\hat{d}_k$ in advance, so that a look-up table can be used for beamforming. Thus, the computational complexity of the numerical optimization is not considered detrimental to the communication performance. 

\begin{remark}
    Note that, since the parameterized precoding structure in \eqref{f_sinc} is based on a sinc-function, $\lvert \left[ \vec{f}(v,\omega)\right]_n\rvert$ can be very small for some antennas, due to the zeros of the sinc-function and its decreasing envelope. As a result, some of the antennas will transmit with very low power and, hence, their impact on the beam is insignificant. As a consequence, the energy consumption that is necessary to operate the antenna array can possibly be reduced by applying a simple threshold-based dynamic antenna selection strategy and thereby reducing the number of active antennas. As an example, assume that the beam parameters are $v = 0.1$ and $\omega=0$ and the ULA consists of $N_\mathrm{t}=64$ elements. Then, if we decide to turn off all antenna elements that are supposed to transmit less than 5\% of the maximum power allocated to a single antenna, only 39 elements would be activated. Thus, in this example, the number of active antennas could be reduced by almost 40\%,
    without significantly affecting the communication performance. 
\end{remark}

 In the following section, we propose a solution to the optimization of $q_k$ in problem \eqref{opt_SO} and suggest an algorithm for the overall tracking and transmission procedure.

\section{Event-based Tracking Algorithm}\label{sec:event+tracking}
\subsection{Event trigger} \label{sec:event}

Recall that $q_k$ is a binary variable that is equal to one when a new channel estimation is performed, and equal to zero otherwise. In other words, we assume to have perfect CSI available at the BS when $q_k=1$ and outdated CSI with a Gaussian distributed user position error if $q_k=0$. Thus, while evaluating the objective function for these two cases is manageable, the challenge for solving problem \eqref{opt_SO} lies in the long-term average overhead constraint \eqref{overhead_const}. To handle this constraint, we use the Lyapunov optimization framework \cite{neelyLyapunov}, which involves defining a virtual queue that indicates the current deviation from the time-average constraint. Subsequently, this virtual queue is stabilized via Lyapunov optimization, which ensures compliance with the long-term constraint. Hence, for our problem, we define a virtual queue $Z$ with $Z_0=0$ as follows:
\begin{equation}\label{Z_dynamic}
    Z_k = \max\{0,~Z_{k-1} + q_k - r_\mathrm{q}\}.
\end{equation}
The corresponding Lyapunov function is given as $L(Z_k) = \frac{1}{2} Z_k^2$.
Thus, the Lyapunov drift is 
\begin{equation}
\begin{aligned}
    \Delta(Z_k) &= \mathbb{E}\left[L(Z_k) - L(Z_{k-1})|Z_k\right]
    \\
    &= \mathbb{E}\left[Z_{k-1}(q_k-r_\mathrm{q}) + \frac{1}{2}(q_k-r_\mathrm{q})^2\right].
    \end{aligned}
\end{equation}

Furthermore, let $G^{(\text{imp})}_{\alpha,k} = \max_{\vec{f}_k} g_\alpha(\vec{f}_k, q_k=0)$ be the optimal value of the objective function given that imperfect CSI with distribution determined by $\sigma_{\mathrm{p},k}$ is available at the BS, whereas $G^{(\text{p})}_{\alpha,k} = \max_{\vec{f}_k} g_\alpha(\vec{f}_k, q_k=1)$ denotes the best achievable value of the objective function under the assumption of perfect channel knowledge. 
Then, the subproblem of optimizing $q_k$ in \eqref{opt_SO} can be formulated as:
\begin{equation}\label{opt_event1}
    \begin{aligned}
        \max_{q_k} \quad & q_k G^{(\text{p})}_{\alpha,k}  + (1-q_k) G^{(\text{imp})}_{\alpha,k} - \mu \Delta(Z_k)\\
        \text{s.t.}\quad 
        & \lim_{K\rightarrow \infty} \frac{1}{K}\sum_{k=1}^{K} q_k \leq r_\mathrm{q}.
    \end{aligned}
\end{equation}
Here, $\mu > 0$ is a predefined weighting parameter. Note that constraint \eqref{power_const} is independent of $q_k$ and can be neglected in this subproblem.
Hence, the solution of \eqref{opt_event1} is obtained by:
\begin{equation} \label{event_condition}
    q_k = \begin{cases} 1 & \text{if}~ Z_{k-1}-r_\mathrm{q}+\frac{1}{2} < \frac{1}{\mu}(G^{(\text{p})}_{\alpha,k}-G^{(\text{imp})}_{\alpha,k})\\
    0 & \text{otherwise.} \end{cases}
\end{equation}
Note that \eqref{event_condition} is based on a dynamic condition that adapts to the system state and hence differs from other state-of-the-art approaches that are based on a fixed threshold.\\
Let $p_{\mathrm{b},k} = P(\gamma_k=0|\gamma_{k-1})$ be the instantaneous blockage probability estimated by the BS at time step $k$ as given in~\eqref{blockage_prob}. Then, with \eqref{R_max}, we have:
\begin{equation*}
    \max_{\vec{f}_k} ~\mathbb{E} [R_k|q_k=1] = (1 - p_{\mathrm{b},k}) R_\mathrm{max}.
\end{equation*} 
Hence, the objective function value with perfect CSI at the BS is obtained by:
\begin{equation}
    G^{(\text{p})}_{\alpha,k} = \begin{cases} \alpha-p_{\mathrm{b},k} & \text{if}~ W \log_2\left(1 + \frac{P_\mathrm{max}\eta^2(\hat{d}_k)}{\sigma_\mathrm{n}^2}\right) \geq R_\mathrm{min}\\
    (1-p_{\mathrm{b},k}) \alpha-(1-\alpha) & \text{otherwise.}
    \end{cases}
\end{equation}
The objective function with imperfect CSI $G^{(\text{imp})}_{\alpha,k}$ is computed based on \eqref{E_rate_vw} and \eqref{P_out_vw} using the optimal beam parameters corresponding to $\sigma_{\tilde{\varepsilon}_k}$ and $d_k$. Note that these can be precalculated and saved in a look-up table along with the corresponding objective function values.

\subsection{Tracking Algorithm}

\begin{figure}[tb]
    \centering
    \scalebox{0.85}{
    \vspace{-.3cm}
    \includegraphics[]{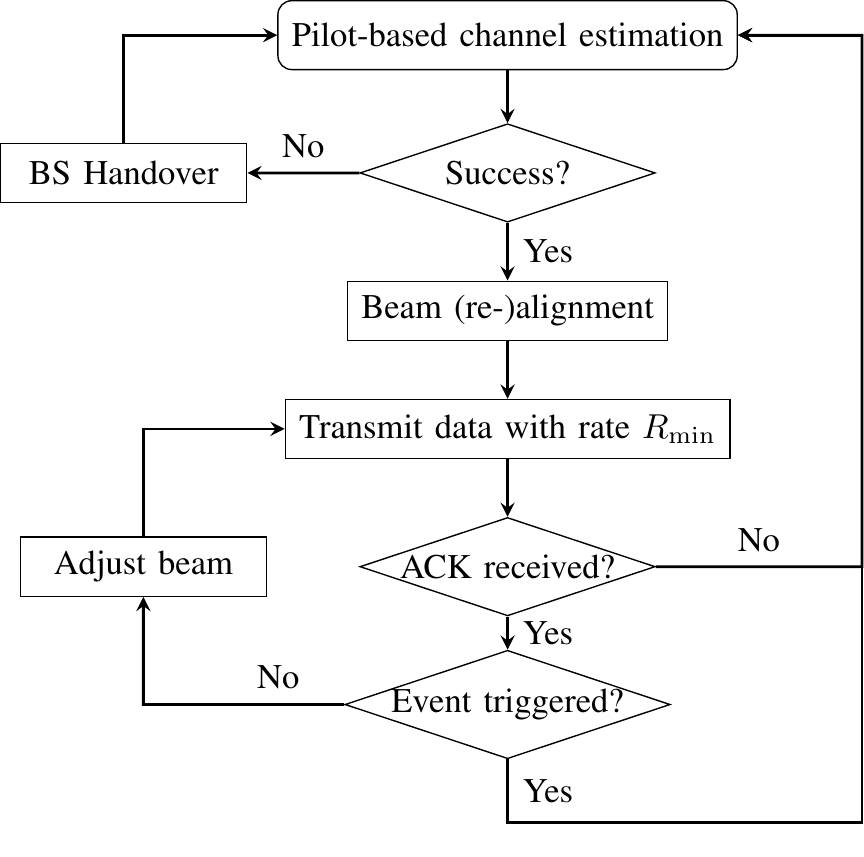}}
    \vspace{-.1cm}
    \caption{Flow diagram showing the tracking algorithm, including event-based pilot transmission and handovers.}
    \label{alg_flow_diagram}
    \vspace{-.2cm}
\end{figure}

Next, we present our framework for the beam tracking and data transmission procedure, including the previously proposed robust beamforming and event-based channel estimation schemes. The dynamic beamwidth adaptation method improves reliable communication despite outdated CSI, i.e., tolerating a higher AoD uncertainty than non-robust beamforming, and therefore enables less frequent channel estimation. However, since the variability of the channel caused by user mobility is not uniform in general, we optimized the time steps at which channel estimation should be performed to ensure timely CSI updates without violating the predefined acceptable amount of overhead on average. However, communication outages can still take place for different reasons, namely dynamic blockage, beam misalignment or exceeding the BS's THz communication range. These events require appropriate reactions, like initiating a new channel estimation, adapting the beamformer or conducting a BS handover. Therefore, in the following, our proposed two-fold scheme is embedded into a communication and tracking procedure.

The overall tracking algorithm is shown in Figure \ref{alg_flow_diagram}. When the considered UE is assigned to a BS, the BS obtains the current position of the UE through pilot signal measurements. After successful channel estimation, beam alignment is performed and data is transmitted with rate $R_\mathrm{min}$. At the end of a time slot, the BS receives a feedback, whether or not decoding was successful at the UE (ACK/NACK signal). We assume that decoding fails only if the actual data rate $R_k$ supported by the channel is less than the transmit data rate $R_\mathrm{min}$. Note that this can happen due to either beam misalignment or blockage, or both. 
If an outage occurs, i.e., the BS received a NACK signal, another pilot signal transmission is invoked to realign the beam. We assume that in case of a blockage event, the channel estimation will fail. In this case, a handover is initiated to assign the UE to a different BS, which is not currently affected by a dynamic blocker. We then switch our perspective to the new serving BS and from then on consider the transmission of the new tagged BS-UE pair.
Otherwise, when the transmitted signal can be successfully decoded by the UE, i.e., the BS received an ACK signal, the event-triggering condition \eqref{event_condition} is checked after each time slot. As long as no pilot transmission event is triggered, the BS will adapt the beamformer and continue to transmit data in the next time slot.

\section{Simulation Results and Analysis}\label{sec:numerical}
The performance of our scheme is now evaluated via simulations. Unless stated otherwise, the main simulation parameters in Table \ref{tab:param} are used. First, we analyze the performance of our adaptive beamwidth precoding scheme based on Monte-Carlo simulations with given AoD uncertainty. After that, the beamforming is embedded into a beam tracking scenario including random walk user mobility for performance evaluation of our proposed event-based tracking approach.

\begin{table}[]
    \centering
        \caption{Parameters used for the simulations, if not stated otherwise.} \vspace{5pt}
    \resizebox{\textwidth}{!}{
    \begin{tabular}{|c|c||c|c|}
    \hline
     Number of ULA antenna elements $N_\mathrm{t}$ ($N_\mathrm{r}$) &  64 (16)
    & Bandwidth $W$ & 10 GHz
    \\ \hline
    Operating Frequency $f$  &  300 GHz
    & Density of dynamic blockers $\lambda_\mathrm{B}$ & 0.3  $\mathrm{m}^{-1}$
    \\ \hline
      Molecular Absorption Coefficient $K(f)$  &  0.0012 $\mathrm{m}^{-1}$
      & Velocity of dynamic blockers  $v_\mathrm{B}$  &  1 m/s
      \\ \hline
      Transmit power $P_\mathrm{max}$  &  30 dBm
      & Height of BS  $h_\mathrm{BS}$  &  3.5 m
      \\ \hline
      Noise power spectral density & -174 dBm/Hz
      & Height of UE   $h_\mathrm{UE}$  &  1.5 m
      \\ \hline
      Time slot duration $T_\mathrm{s}$  &  50 ms
      & Height of blockers    $h_\mathrm{B}$  &  1.8 m
      \\ \hline
      RW step size standard deviation $\sigma_\mathrm{m}$  & 0.05 m 
      & Unblocking rate    $\mu_\mathrm{B}$  &  3 $\mathrm{s}^{-1}$
      \\ \hline 
    \end{tabular}} 
    \vspace{-.5cm}
    \label{tab:param}
\end{table}

\subsection{Beamforming Scheme}
\begin{figure}[t]
    \centering
     \subfigure[Pareto region]{\label{fig:pareto_boundary}
    \includegraphics[]{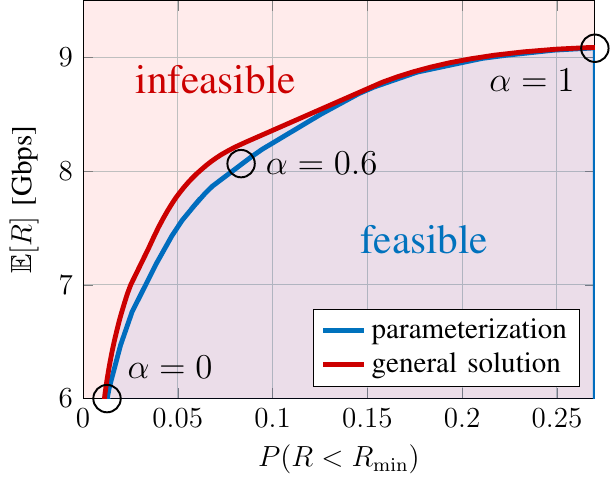}}
    \subfigure[Beam pattern]{\label{fig:beam_pattern}
    \includegraphics[]{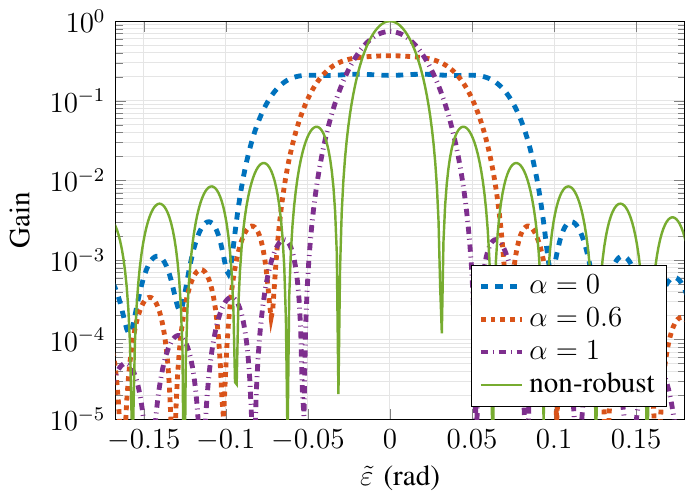}}
    \caption{\subref{fig:pareto_boundary} Pareto boundary and feasible region of the generally optimized beamformer and the achievable region of the proposed parameterized beamformer, with $R_\mathrm{min}=5$ Gbps, $d=8$ m, and $\sigma_{\tilde{\varepsilon}}=1.5^\circ$. \subref{fig:beam_pattern} Beam pattern with optimized parameters corresponding to the three points marked in \subref{fig:pareto_boundary} in comparison to the non-robust beam.}
    \label{fig:pareto}
\end{figure}
We first examine the performance of our variable-beamwidth precoding scheme proposed in Section \ref{sec:beamformer_opt}.
We identify the Pareto boundary by numerically solving the optimization problem \eqref{opt_BF} with a general precoder as described in Remark \ref{remark_general_sol} for varying weight parameter $\alpha$. In Figure \ref{fig:pareto_boundary}, we compare the achievable region of our parameterized beamformer with the general Pareto region for a BS-UE distance of $d=8$ m and AoD standard deviation $\sigma_{\tilde{\varepsilon}}= 1.5^\circ$. Recall that $\alpha=0$ corresponds to minimizing the outage probability, whereas $\alpha=1$ is maximizing the expected rate. In fact, the maximum expected rate and minimum outage probability are opposing objectives, i.e., increasing the rate expectation comes at the cost of a higher outage probability, while reducing outages entails a loss in expected rate. Each of the two objectives alone lead to substantially different beamforming strategies. When minimizing the outage probability ($\alpha=0$), the expected rate reduces by one third. Maximizing the expected rate ($\alpha=1$) leads to an increase of the outage probability by a factor of more than 20. Note that when $\alpha$ is close to one, we have to accept a much higher outage probability for a relatively small gain in terms of expected rate. The opposite effect is observed when $\alpha$ is close to zero. This motivates considering a multi-objective optimization problem in order to  balance the two objectives.

Although our proposed parameterized approach does not fully achieve the Pareto region of a general beamformer, it is shown to clearly be a useful approximation despite its complexity being much lower. Especially for higher values of $\alpha$ (i.e., higher weighting of the expected rate), the gap between the parameterized and the general solution is negligible. The biggest performance gap is observed in the area around $\alpha=0.5$ when both objectives are balanced. In particular, the parameterization results in an average rate loss of about 0.27 Gbps at most, which corresponds to a rate reduction of approximately 3.4\%, whereas the outage probability increases by up to 0.018, i.e., by around 25\% at most. Hence, the performance loss of the parameterized beamformer mostly stems from the outage probability minimization. 

In Figure \ref{fig:beam_pattern}, the beam pattern of the optimized parameterized beams is shown for three cases, namely $\alpha \in \{0,0.6,1\}$ in comparison to the non-robust beam. As we can see, the beam gets wider as we take greater account of the outage probability, i.e., as $\alpha$ decreases. However, even for $\alpha=1$ the chosen beam is wider than the non-robust beam. Besides that, we notice that the beam also gets flatter when $\alpha$ decreases. The reason for that effect lies in the threshold-based definition of outage probability, meaning that outages are reduced when the beam gain is above a threshold for most channel realizations, while the actual value of the gain is less relevant.

\begin{figure}[t]
    \centering
    \subfigure[Expected rate maximization $(\alpha=1)$]{\label{fig:v_rate_contour}
   \includegraphics[]{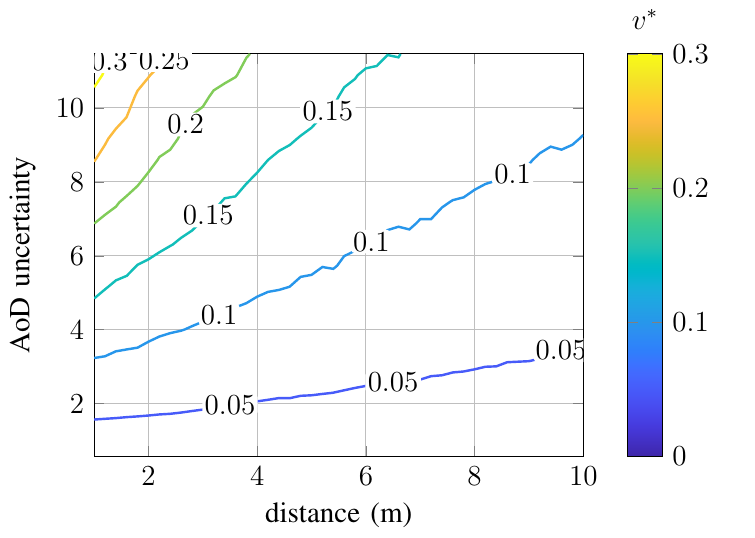}}
    \subfigure[Outage Probability minimzation $(\alpha=0)$]{\label{fig:v_out_contour}
    \includegraphics[]{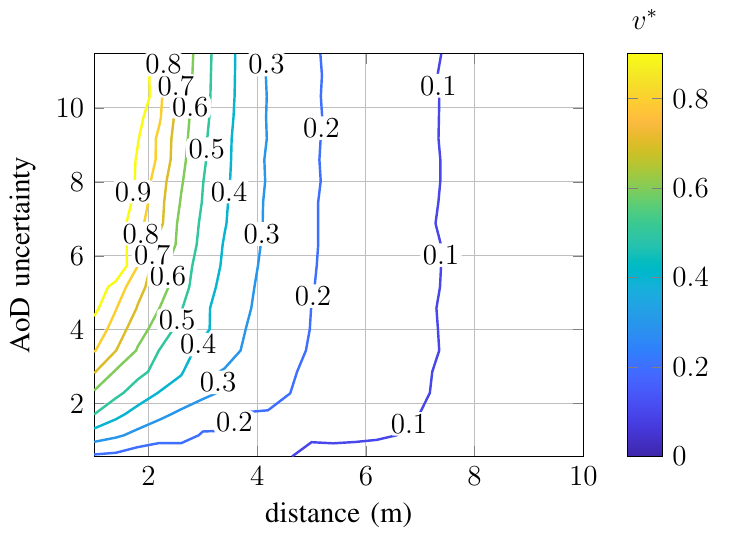}}
    \caption{Contour plot of the optimized beamwidth parameter $v$ as a function of distance and AoD deviation $\sigma_{\tilde{\varepsilon}}$, for $\alpha=1$ and $\alpha=0$. Larger values of $v$ lead to a wider beam, while $v=0$ corresponds to the non-robust beamformer.}
    \label{fig:contour}
\end{figure}
 
Figure \ref{fig:contour} shows the optimal beamwidth parameter $v$ that solves \eqref{opt_BF_vw} for different BS-UE distances $d$ and angular deviation $\sigma_{\tilde{\varepsilon}}$, for the two marginal cases $\alpha=1$ (Fig. \ref{fig:v_rate_contour}) and $\alpha=0$ (Fig. \ref{fig:v_out_contour}). Since $v$ is directly related to the beamwidth, we gain insights on how the optimal beamwidth changes in different scenarios. While $v=0$ corresponds to the non-robust beamformer, i.e., a narrow beam, a larger value of $v$ leads to a wider beam. Hence, from Figure \ref{fig:v_rate_contour}, we observe that when maximizing the expected rate, the most prominent factor leading to a wider beam is a higher $\sigma_{\tilde{\varepsilon}}$. Indeed, when the position of the UE is subject to more fluctuation a wider beam is necessary to cover any prospective and sudden changes in the UE's position. Additionally, the figure also shows that when the distance is small, relying on beamforming to concentrate the power and compensate for the THz propagation loss is not as necessary as for longer distances, i.e., a wide beam is more beneficial to increase robustness when the user is sufficiently close to the BS. For instance, with an AoD standard deviation of $\sigma_{\tilde{\varepsilon}} = 8^\circ$ the optimal beamwidth parameter is 0.1 if the UE is at 8~m distance, but increases to 0.2 if the distance is only 2~m. Clearly, beamforming is inevitable when the power needs to be sustained for a longer range at THz frequency bands.
Intuitively, this represents the tradeoff between increasing the probability of coverage with a wider beam when the user's position is uncertain and increasing directivity to enhance the received signal strength when facing severe path loss in the THz band.

In Figure \ref{fig:v_out_contour}, when minimizing the outage probability, we again observe an increase in beamwidth for higher AoD uncertainty and a decreasing beamwidth for higher communication distance. However, it is clear that both objectives require significantly different beamforming strategies. More specifically, for the most part, outage probability minimization leads to substantially wider beams than expected rate maximization, especially for lower communication distances (below 5m). When the channel gain is sufficiently high, the transmission power can be spread more widely without causing an outage and hence, the outage probability is reduced. However, a higher gain in the directions that are most likely is beneficial when considering the expected rate, hence, a moderate beamwidth is preferred in this case. In addition to that, Figure \ref{fig:v_out_contour} shows that for higher AoD uncertainty, the beamwidth depends almost exclusively on the communication distance, i.e., the contour lines become nearly vertical. For instance, at a distance of $4$ m the optimal value for $v$ first increases with growing AoD uncertainty, but remains at a value close to 0.3 for $\sigma_{\tilde{\varepsilon}}>6^\circ$. The intuition behind that follows from the fact that as soon as the AoD uncertainty becomes detrimental to the outage probability, making the beam as wide as possible is beneficial. Here, despite an increase in the AoD uncertainty, the beamwidth cannot be further increased. Meanwhile, when the communication distance is higher, narrowing the beams is necessary to prevent outages due to the high path loss in the THz band.

\begin{figure}[]
    \centering
    \scalebox{0.7}{
    \includegraphics[]{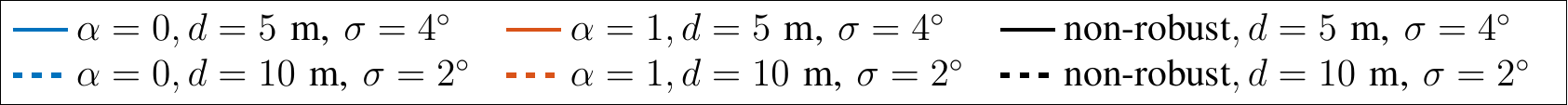}}\\ \vspace{-2mm}
    \subfigure[Optimized beamwidth]{\label{fig:beamwidth_mol_abs}
   \includegraphics[]{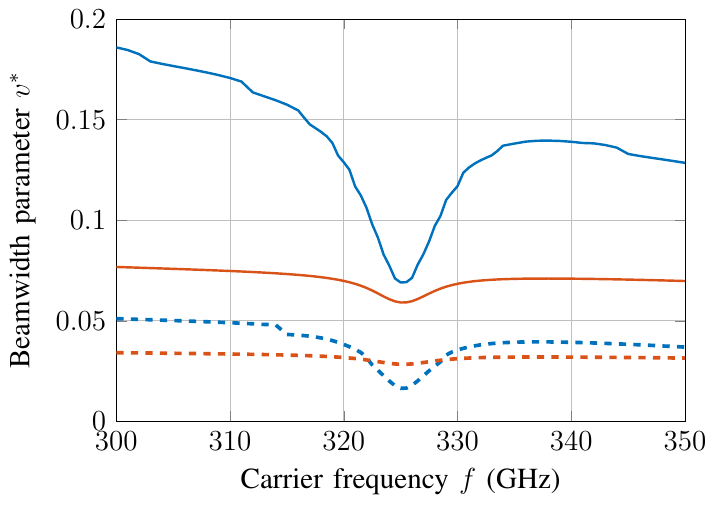}}
    \subfigure[Outage probability]{\label{fig:outage_mol_abs}
    \includegraphics[]{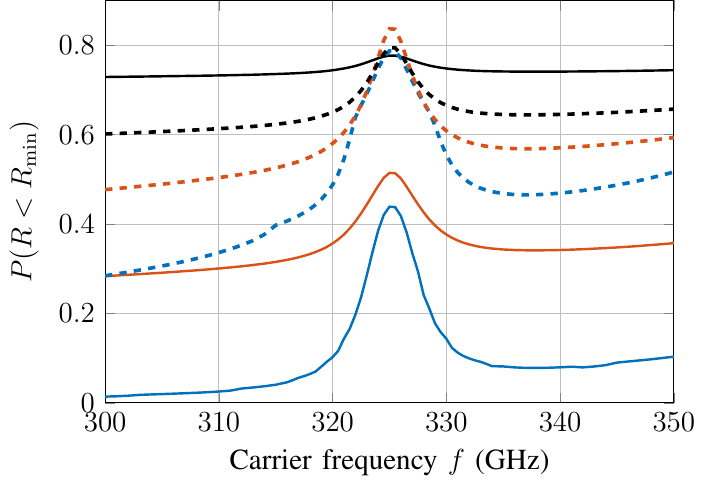}}
    \caption{Impact of molecular absorption on the optimal beamwidth parameter and outage probability with $R_\mathrm{min}=5$ Gbps, shown for outage probability minimization (blue curves) and expected rate maximization (red curves). For both transmission distances of 5 m (solid lines) and 10 m (dashed lines), the misalignment standard deviation is 35 cm, which is equivalent to $4^\circ$ and $2^\circ$ in terms of AoD deviation, respectively.}
    \label{fig:mol_abs}
\end{figure}

Figure \ref{fig:mol_abs} showcases the effect of molecular absorption on our proposed beamforming scheme. The optimal beamwidth for maximizing expected rate and minimizing outage probability are shown for the frequency range $[300~\text{GHz}, 350~\text{GHz}]$ in Fig. \ref{fig:beamwidth_mol_abs}. Note that there is an absorption line at around 325 GHz caused by the absorption of the water molecules \cite{kokkoniemi2021line}. Here, the frequency band surrounding the absorption line necessitates a much narrower beam to compensate for three THz-specific factors, namely, \emph{the THz space path loss, the molecular absorption effect, and re-radiation noise}. As a result, the beamforming strategy differs substantially for the two objectives. While the absorption line clearly has an impact in both cases, it is much more pronounced when the outage probability is considered as our objective. Since all channel realizations are affected equally by molecular absorption, a small beam adjustment is sufficient when considering expected rate, while the beamwidth parameter is more than halved when optimizing the outage probability in order to meet the rate requirement within the main lobe.
Furthermore, with good channel conditions, the outage probability objective benefits from a wider beam to prevent misalignment. However, in contrast to mmWave, the channel is heavily affected by molecular absorption. Here, the severe path loss becomes the main reason for outages (e.g. at $d=10$ m and $f=325$ GHz). Hence, narrowing the beam (even more than in the expected rate-focused scheme) becomes a preferred strategy with respect to outage probability.
Figure \ref{fig:outage_mol_abs} shows the probability of the data rate dropping below the target rate of 5 Gbps. Since the outage probability is clearly affected by molecular absorption, optimizing the beamwidth is essential, especially for smaller communication distances, which are common in THz systems, and where the same movement leads to higher AoD deviation. Note that maximizing the expected rate can still lead to many outages caused by misalignment. Furthermore, at a transmission distance of 10 m, we observe that even relatively small changes of the beamwidth can significantly impact the outage probability.

 \begin{figure}[]
    \centering
    \subfigure[Expected rate]{\label{fig:rate_bs}
   \includegraphics[]{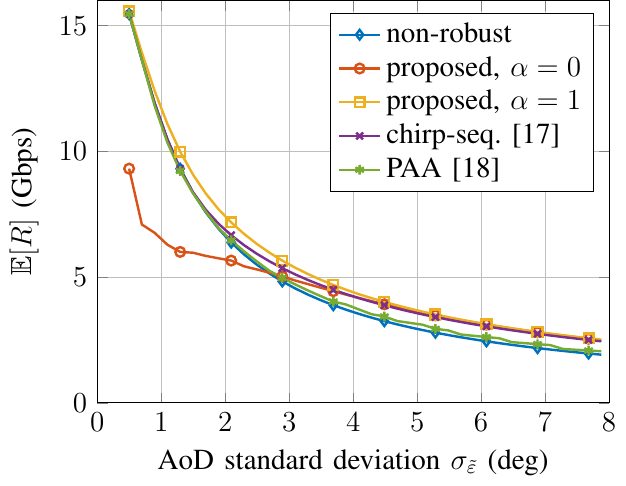}}
    \subfigure[Outage probability]{\label{fig:outage_bs}
    \includegraphics[]{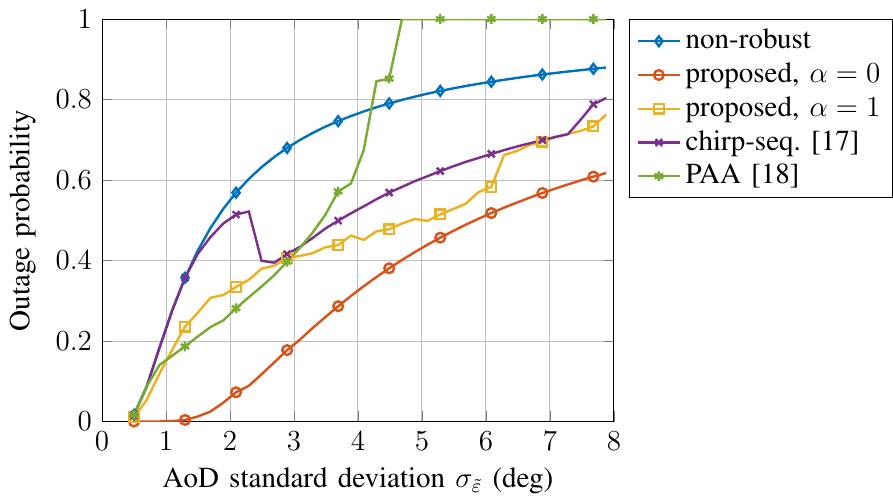}}
    \caption{Expected rate and outage probability as a function of the AoD standard deviation for a communication distance of $d=8$ m and target rate $R_\mathrm{min} = 5$ Gbps. The curves show our proposed beamforming scheme for the marginal cases $\alpha=0$ and $\alpha=1$, compared to non-robust beamforming and two baseline schemes.}
    \label{fig:baseline_comp}
\end{figure}

Figure \ref{fig:baseline_comp} compares our proposed parameterized beamformer to non-robust beamforming and the following two variable-beamwidth benchmark schemes proposed for mmWave systems:
 
 \paragraph{Chirp-sequence-based Beamformer \cite{peng2017robustWB}}
The authors in \cite{peng2017robustWB} proposed a beamforming scheme based on Zadoff-Chu-sequences with a parameter, that adjusts the beamwidth and is numerically calculated so that the expected data rate is maximized.
In order to modify the shape of the beam, the authors suggest to additionally apply a triangular window to the precoder. However, they do not propose a strategy on how to optimally select the window. Therefore, we omit the use of windowing when comparing this scheme to our proposed approach.

\paragraph{Partial Antenna Array Activation \cite{chung2020adaptiveBT}}
The authors in \cite{chung2020adaptiveBT} suggest to only activate part of the antenna array in order to form a wider beam. The number of active antennas is determined based on a heuristic, so that the half-power beamwidth approximately covers the range $[\hat{\varphi}-\sigma_\varepsilon, \hat{\varphi}+\sigma_\varepsilon]$.
Note that different from our scheme and the one from \cite{peng2017robustWB}, the beamwidth is completely independent of the SNR and path loss.

We analyze the expected rate and the outage probability as a function of the AoD standard deviation for a fixed communication distance of $d=8$ m in the Figures \ref{fig:rate_bs} and \ref{fig:outage_bs}, respectively. 
While the expected rate decreases with growing AoD uncertainty for all schemes, Figure \ref{fig:rate_bs} demonstrates the superiority of our proposed approach with $\alpha=1$ in terms of the expected rate for all $\sigma_{\tilde{\varepsilon}}$. Note that our scheme with $\alpha=0$, where outage probability is the only objective considered, achieves a much lower average rate than all baselines for $\sigma_{\tilde{\varepsilon}}<2.5^\circ$, but converges to the expected rate achieved by the scheme with $\alpha=1$ as $\sigma_{\tilde{\varepsilon}}$ increases. The chirp-sequence based approach from \cite{peng2017robustWB}, which also aims at maximizing the expected rate, performes similar to the non-robust beamformer when $\sigma_{\tilde{\varepsilon}}$ is low, and only converges to the rate achieved by our proposed scheme for higher AoD uncertainty ($\sigma_{\tilde{\varepsilon}}>4^\circ$). The partial antenna activation scheme \cite{chung2020adaptiveBT} is based on a heuristic and neither maximizing expected rate nor minimizing outage probability explicitly. In terms of rate expectation, it is shown to be just slightly better than the non-robust beamforming, with a gap of around 0.6 Gbps to our proposed scheme.
Figure \ref{fig:outage_bs} compares the outage probability of the respective beamforming approaches. Here, our proposed scheme with $\alpha=0$, in which the outage probability is minimized, proves to be superior to all baselines for all $\sigma_{\tilde{\varepsilon}}$. Most significantly, for $\sigma_{\tilde{\varepsilon}}<1.5^\circ$, it has a considerably smaller slope than all other schemes, and up to $\sigma_{\tilde{\varepsilon}}=3^\circ$, the gap in outage probability is around 0.2. Note that none of the benchmark schemes are designed to minimize outage probability.

\subsection{Beam Tracking Simulation}

\begin{figure}[]
    \centering
    \subfigure[Time sequence example]{\label{fig:event_example}
   \includegraphics[]{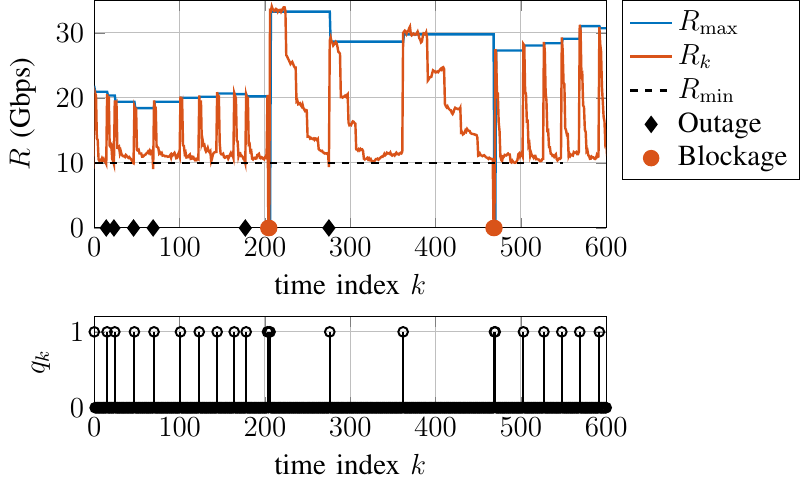}}
    \subfigure[CDF of achievable rate]{\label{fig:CDF_tracking}
    \includegraphics[]{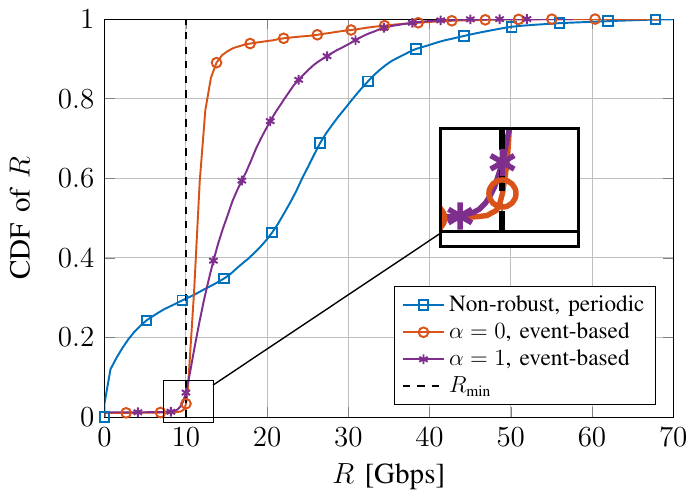}}
    \caption{Illustration of the event-based tracking procedure with $r_\mathrm{q} = 0.05$ and target rate $R_\mathrm{min}=10$ Gbps: \subref{fig:event_example} Example of achievable data rate over time with corresponding pilot transmission event times, \subref{fig:CDF_tracking} CDF of the data rates achieved by our proposed scheme compared to a non-robust baseline with periodic pilot transmission.}
    \label{fig:event_scheme}
\end{figure}

Next, we simulate the beam tracking scenario considering a random walk of the UE and parameters in Table \ref{tab:param}.
In Figure \ref{fig:event_scheme}, we analyze the achievable rates when using our proposed beamforming scheme combined with the proposed event-based tracking scheme according to \eqref{event_condition} with $\mu=0.5$ and average overhead limitation $r_\mathrm{q}=0.05$, while the target rate is set to $R_\mathrm{min} = 10$ Gbps. Figure \ref{fig:event_example} illustrates the event-triggering behaviour for a period of 600 time steps. The upper graph shows the rate with perfect CSI as expected by the BS, namely $R_\mathrm{max}$, as well as the actual achievable rate with the selected beamformer (as in Section \ref{sec:sinc_BF} with $\alpha=0.6$), denoted by $R_k$. Note that $R_\mathrm{max}$ stays constant in between channel estimation events. The bottom graph shows the corresponding event times given by $q_k$.
While a nearly periodic pilot pattern can be observed for certain time spans (e.g. $k \in [500,600]$), the channel estimation events occur in a non-uniform manner in general. In particular, when an outage occurs (represented by a black diamond shape), i.e., $R_k$ drops below the target rate $R_\mathrm{min}$, a new channel estimation is initiated. A handover is performed to handle blockages (depicted as a red circle). In the evaluation, we pick a serving BS at a random position in the range of 3 to 7 meters apart from the user to simulate a handover. Note that these immediate reactions to outage and blockage events lead to an increase in overhead, which can be observed, e.g., $k \in [0,70]$, as well as around $k=170$ and $k=205$. In the subsequent time steps, the interval between pilot transmission events is increased to compensate for the excess overhead. Furthermore, the user's relative position to the BS has an impact on the frequency of pilot events as well. This phenomenon can be observed at $k=205$, where a blockage event induces a handover to a BS that is better placed and allows for much longer intervals between pilots. This event-based scheme makes the data transmission more efficient, since the system can prevent outages resulting from beam misalignment by performing regular channel estimations, while still being able to immediately react to outage and blockage events without violating the average overhead constraint in the long term.

Next, \ref{fig:CDF_tracking} shows the cumulative distribution function (CDF) of the data rates achieved by the proposed beamforming and tracking scheme for the two special cases $\alpha=0$ and $\alpha=1$, compared with a basic non-robust beamformer with periodic pilot measurements.
The CDF of our proposed event-based approach with $\alpha=1$ is below the non-robust CDF for low data rates (below 13 Gbps). This is a result of the fact that our variable beamwidth precoder exhibits a smaller prospect for a low data rate, which is reduced even further by the event-based tracking approach. Meanwhile, the CDF of our proposed event-based scheme grows above the non-robust CDF beyond 13 Gbps since very high data rates (higher than 30 Gbps) are also less likely. 
For our proposed scheme with $\alpha=0$, which minimizes the outage probability, the CDF is the lowest for data rates below the target rate $R_\mathrm{min}$, but then rapidly grows above the other CDFs. When this scheme is applied, most data rates lie between 10 and 15 Gbps. 
Although the non-robust baseline enables more high data rates (above 30 Gbps) than our proposed schemes, there are also much more low rates in this case. More precisely, around $30\%$ of the rates are below the target rate of 10 Gbps with the non-robust scheme, while this is the case for only $3\%$ ($\alpha=0$) and $6\%$ ($\alpha=1$) of the rates achieved with the other two schemes.

\begin{figure}[t]
    \centering
    \includegraphics[]{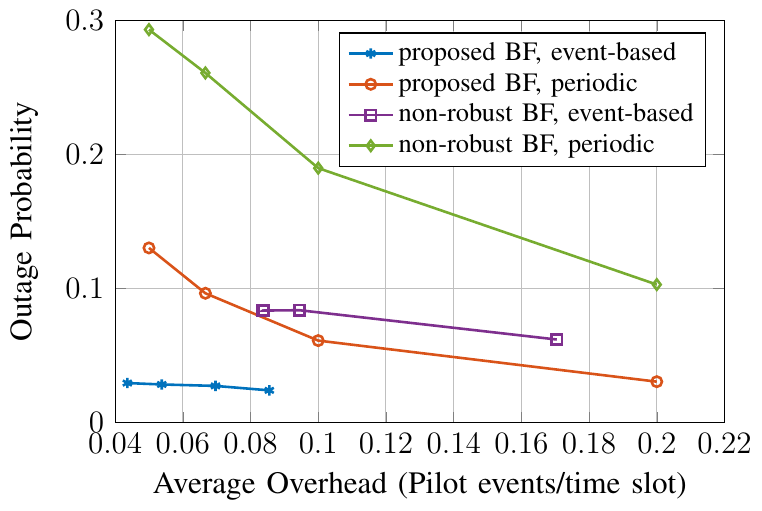}
    \caption{Outage probability as a function of average overhead for a beam tracking simulation over 100000 time steps, with target rate $R_\mathrm{min}=10$ Gbps, $\alpha=0.6$ and an overhead constraint with $r_\mathrm{q} \in \{0.05,~0.667, ~0.1, ~0.2\}$.}
    \label{fig:outage_overhead}
\end{figure}

Since we are interested in reliable communication, we study the relation between the frequency of outage events and pilot transmission overhead in Figure \ref{fig:outage_overhead} and study the efficiency of our event-based channel estimation scheme compared to periodic pilot transmission. 
It is demonstrated, that although our proposed beamformer as well as the event-based tracking approach, when used individually, can significantly reduce the amount of outages compared to non-robust periodic tracking, a combination of both proposed schemes enables much more reliable and efficient communication.
With periodic channel estimation, our proposed beamforming scheme with $\alpha=0.6$ reduces the amount of outages by more than $50\%$ compared to non-robust beamforming. 
When applying the proposed event-triggered scheme, the actual average overhead can differ from the selected $r_\mathrm{q}$. In fact, with the event-based, but non-robust scheme the average overhead is at least 0.083, i.e., there is no feasible solution when $r_\mathrm{q}$ is below this value. This is because the tracking procedure in Figure \ref{alg_flow_diagram} enforces pilot signal transmission following each outage event regardless of the overhead constraint. 
With the combination of the proposed precoding and event-based tracking scheme, however, the average overhead is far below $r_\mathrm{q}$. Moreover, the outage probability is substantially lower than in all other cases, namely below $3\%$ for all $r_\mathrm{q}$. Indeed, an outage probability slightly below $3\%$ is achieved by the proposed combined scheme with one pilot event every 23 time steps on average, while the proposed BF with periodic channel estimation requires one pilot event every 5 time steps to achieve the same. 
Hence, we prove that the combination of our proposed beamwidth adaptation approach and an event-triggered tracking scheme significantly improves communication reliability, while requiring substantially less overhead on average.

\section{Conclusion}\label{sec:conc}
In this paper, we have proposed a reliable low-overhead communication scheme for a beam tracking scenario in the THz frequency band. Given the adoption of narrow pencil beams at THz communication links, beam misalignment is a fundamental challenge for mobile users that needs to be addressed. Consequently, in this work, we scrutinize the tradeoff between increasing coverage probability and supporting a considerable communication range at THz frequencies. In particular, we have formulated a multi-objective optimization problem that maximizes the expected data rate and minimizes the outage probability. We have proposed a dynamic beamwidth adaptation scheme based on a parameterized precoder. In order to maintain a low channel estimation overhead, an event-based tracking scheme has been presented, which dynamically adjusts the pilot transmission intervals in the presence of user mobility and dynamic blockage.
Simulation results show that our proposed precoder outperforms state of the art adjustable beamwidth approaches. Our scheme has been shown to significantly reduce the amount of communication outages without violating restrictions on the average overhead.

\begin{spacing}{1.45}
\bibliographystyle{IEEEtran}
\bibliography{references}
 \end{spacing}

\end{document}